\documentclass[12pt]{iopart}
\usepackage{iopams}
\usepackage{graphicx}

\newcommand{\ve}{\varepsilon}
\newcommand{\veh}{\varepsilon_h}%{\text{SiO}_2}}
\newcommand{\vem}{\varepsilon_{\rm{Ag}}^{\rm{NL}}}
\newcommand{\om}{\omega}

\begin{document}

\title{Nonlinear metal-dielectric nanoantennas for light switching and routing}

\author{R. E. Noskov,$^{1}$ A. E. Krasnok$^1$ and Yu. S. Kivshar$^{1,2}$}

\address{$^{1}$National Research University of Information Technologies, Mechanics and Optics (ITMO), St. Petersburg 197101, Russia \\
$^{2}$Nonlinear Physics Centre, Research School of Physics and Engineering, Australian National University, Canberra ACT 0200, Australia}
\ead{nanometa@gmail.com}

\begin{abstract}
We introduce a novel hybrid metal-dielectric nanoantenna composed of  dielectric (crystalline silicon) and metal (silver) nanoparticles. A high-permittivity dielectric nanoparticle allows to achieve effective light harvesting, and nonlinearity of a metal nanoparticle controls the radiation direction. We show that the radiation pattern of such a nanoantenna can be switched between the forward and backward directions by varying only the light intensity around the level of 11 MW/cm$^2$, with the characteristic switching time of 260 fs.
\end{abstract}

\maketitle

\section{Introduction}

The study of optical nanoantennas became a subject of intensive research~\cite{Bharadwaj,Novotny}. Nanoantennas hold a promise for subwavelength manipulation and control of optical radiation for solar cells and sensing~\cite{Maier}. In a majority of applications, a crucial factor is a control over nanoantenna radiation pattern and its tunability. Spectral tunability and variable directionality have been proposed for bimetallic antennas~\cite{Alivisatos,Shegai}, the Yagi-Uda architectures~\cite{Kosako,Dorfmuller}, mechanically reconfigurable Au nanodimers~\cite{Huang}, high-permittivity dielectric nanoparticles~\cite{Devilez,Krasnok}, and for nanoparticle chains~\cite{Koenderink}. In addition, several suggestions employed the concept of plasmonic nanoantennas with a nonlinear load where the spectral tunability is achieved by varying the pumping energy~\cite{Alu,Abb,Maksymov}.

\begin{figure}[b]
\centerline{\mbox{\resizebox{7cm}{!}{\includegraphics{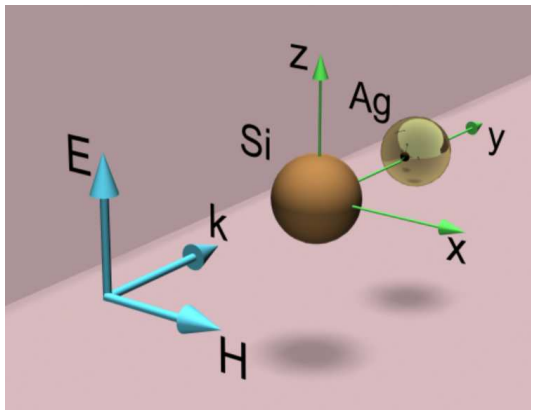}}}}
\caption{\label{fig:1} Schematic view of a dimer metal-dielectric nanoantenna.}
\end{figure}

In this paper, we suggest and study theoretically a novel type of metal-dielectric nanoantenna structures composed of a pair of dielectric (e.g., crystalline silicon) and metal (e.g., silver) nanoparticles. The use of a high-permittivity dielectric nanoparticle allows achieving the efficient light concentration, whereas the nonlinear response of a metal nanoparticle helps to control the radiation direction of the nanoantenna. As a result, for such a structure we can realize the efficient dynamical control over the scattering pattern by varying the external field intensity. An estimated switching time of 260 fs along with relatively low required intensities of $11$~MW/cm$^2$ opens a promising perspective for using nonlinear metal-dielectric nanoantennas in logical and switching devices.

\section{Model}

We consider a pair of spherical silicon and silver nanoparticles embedded into a SiO$_2$ host medium with permittivity $\veh$, excited by a plane wave, as shown in Fig.~\ref{fig:1}. We assume that the radii of the metallic and dielectric nanoparticles and the center-to-center distance are $R_{\rm{Ag}}=15$~nm, $R_{\rm{Si}}=30$~nm, and $d=80$~nm, respectively. Ratio $R_{\rm{Ag}}/d$ satisfies the condition $R_{\rm{Ag}}/d \leq1/3$, so that we can employ the point dipole approximation~\cite{Khlebtsov}. In the optical spectral range, a linear part of a silver dielectric constant can be written in a generalized Drude form $\ve_{\rm{Ag}}^{\rm{L}}=\ve_{\infty}-\om_p^2/[\om(\om+i\nu)]$, where $\ve_\infty=4.96$, $\hbar\om_p=9.54$~eV, $\hbar\nu=0.055$~eV~\cite{Johnson} [$\exp(-i\om t)$ time dependence is assumed]; whereas dispersion of SiO$_2$ can be neglected since $\veh\simeq2.2$ for photon energies 2.8 –- 3.2~eV~\cite{Palik}. In this range a permittivity of silicon $\ve_{\rm{Si}}$ changes approximately from 23 to 36 \cite{Palik}, and we take this experimental data into account but below it will be shown that the impact of silicon dispersion on the system dynamics is insignificant as well. A nonlinear dielectric constant of silver is $\ve_{\rm{Ag}}^{\rm{NL}}=\ve_{\rm{Ag}}^{\rm{L}}+\chi^{(3)}|{\bf E}^{(\rm{in})}|^2$, where ${\bf E}^{(\rm{in})}$ is the local field inside the particle. We keep only cubic susceptibility due to spherical symmetry of the silver particle. According to the model suggested in Ref.~\cite{Drachev}, 15 nm radii Ag spheres possess a remarkably high and purely real cubic susceptibility $\chi^{(3)}\simeq 6\times 10^{-9}$~esu, in comparing to which the cubic nonlinearity of both Si and SiO$_2$ are negligibly weak ($\thicksim 10^{-12}$~esu~\cite{Dinu} and $\thicksim 10^{-15}$~esu~\cite{weber_book_03}, respectively).

We have chosen silver among other metals due to its relatively low loss as well as high nonlinear susceptibility, while quite a high permittivity of silicon leads to pretty large extinction cross-section of the Si nanoparticle that makes possible to achieve the directional scattering from an asymmetric dimer. Fused silica was taken as a host matrix to shift the surface plasmon resonance frequency of the Ag nanoparticle to blue light where silicon shows relatively small losses. In addition, SiO$_2$ possesses good enough optical transparency \cite{Palik} to maintain strong laser powers needed for observation of nonlinear switching.

First, we start from the Fourier transforms of the particle electric dipole moments
\begin{equation}\label{Fourier transforms}
\begin{array}{lcl}
p_{{\rm Si,}z}=\alpha_{\rm{Si}}( E_z^{(\rm{ex})}+A p_{{\rm Ag,}z}),\\
p_{{\rm Ag,}z}=\alpha_{\rm{Ag}}( E_z^{(\rm{ex})}\exp(ikd)+A p_{{\rm Si,}z}),
\end{array}
\end{equation}
where $A=[\exp(ikd)/(\veh d)][k^2-(1/d^2)+ (ik/d)]$ describes the dipole-dipole interaction between particles, $k=\om/c\sqrt{\veh}$ is the wavenumber, $c$ is the speed of light, $E_z^{(\rm{ex})}$ is the amplitude of the plane wave, $$\alpha_{\rm{Ag}}=\veh\left\{\frac{\vem(\om)+2\veh}{R_{\rm{Ag}}^3[\vem(\om)-\veh]}+i \frac{2}{3} k^3\right\}^{-1}$$ and $\alpha_{\rm Si}=3i a_1/(2k^3)$ are polarisabilities of silver and silicon particles, $a_1$ is the electric-dipole Mie scattering coefficient \cite{Stratton}. Small size of the silver nanoparticle allows us to use a quasistatic expression for $\alpha_{\rm{Ag}}$ with the radiation damping correction. On the other hand, the local field inside a 60 nm silicon particle with quite a large permittivity can not be considered as purely homogenous. That is why we define $\alpha_{\rm{Si}}$ through the electric-dipole Mie scattering coefficient.
\begin{figure}[t]
\centerline{\rm{\resizebox{9cm}{!}{\includegraphics{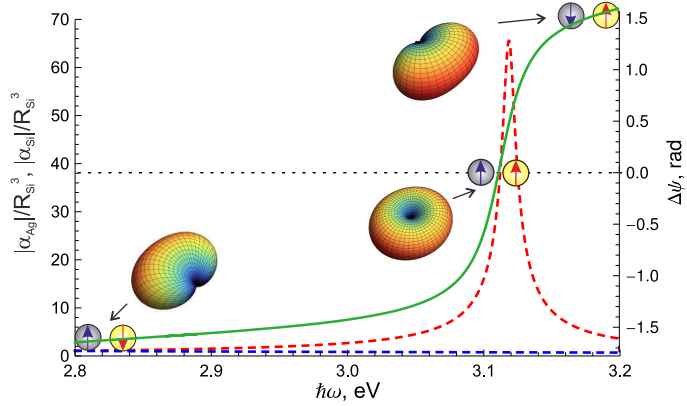}}}}
\caption{\label{fig:2} The full green line denotes relative phase shift between dipole moments of nanoparticles, while dashed lines show polarizability modulus of silver (red) and silicon (blue) nanoparticles normalized to $R_{\rm Si}^3$ versus frequency. Red and blue arrows indicate the orientation of the dipole moments induced in the particles at different frequencies. Insets show scattering patterns of a nanodimer when: 1) $\hbar\om=2.8$~eV -- the system scatters light in the backward direction; 2) $\hbar\om=3.14$~eV -- the scattering pattern is omnidirectional; 3) $\hbar\om=3.2$~eV -- the system scatters light in the forward direction.}
\end{figure}

To illustrate how to achieve the directional scattering from an asymmetric dimer, we analyze the scattering intensity of two dipoles which
is given by
\fl
\begin{eqnarray*}
U(\phi,\theta)=\frac{ck^4}{8\pi \veh^{3/2}} \sin^2 \theta & \left[ |p_{{\rm Si,}z}|^2+ |p_{{\rm Ag,}z}|^2 \right. \\
& + \left. 2 |p_{{\rm Si,}z}| |p_{{\rm Ag,}z}| \cos(\Delta\Psi+kd \sin \theta \sin \phi) \right],
\end{eqnarray*}
where $\phi$ and $\theta$ are spherical azimuthal and polar angles, respectively, and $\Delta\Psi$ denotes an internal phase shift between the two dipoles. $\Delta\Psi$ is determined by the complex particle polarisabilities and therefore varies with the size, shape and material composition  of  the  particles.  The first two terms in this expression describe the individual dipole contributions; while the latter is responsible for interference between particle fields. Clearly, directional scattering can be obtained when $\Delta\Psi$ sufficiently differs from zero. In the linear limit  $\Delta\Psi$ can be tuned in a wide range by variation in the frequency since the silver particle experiences the strong surface plasmon resonance at $\hbar\om_0=\hbar\om_p/\sqrt{\ve_\infty+2\veh}=3.14$~eV; whereas $\alpha_{\rm{Si}}$ is almost frequency independent, as shown in  Figure  \ref{fig:2}. Similar to other nanoparticle systems with broken symmetry \cite{Alivisatos,Shegai,Devilez,Krasnok}, one may see switching the dimer scattering pattern during the growth of $\om$.

\section{Results and discussions}

Next, we study nonlinear dynamics of the dimer by employing the dispersion relation method~\cite{Whitham} that allows to derive a system of the coupled equations for slowly varying amplitudes of the particle dipole moments. This approach is based on the assumption that in the system there exist small and large time scales, which, in our case, is fulfilled automatically because the silver particle acts as a resonantly excited oscillator with slow (in comparison with the light period) inertial response; whereas the almost frequency independent $\alpha_{\rm{Si}}$ allows us to treat the silicon particle response as instantaneous.
\begin{figure}[t]
\centerline{\rm{\resizebox{13cm}{!}{\includegraphics{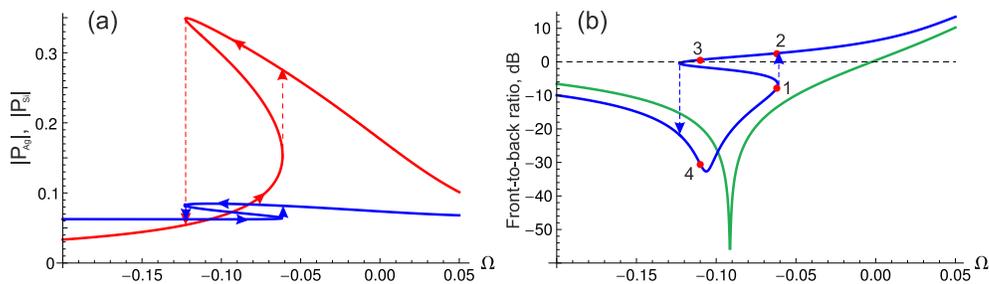}}}}
\caption{\label{fig:3} Frequency dependencies of (a) $|P_{\rm{Ag}}|$ (red), $|P_{\rm{Si}}|$ (blue) and (b) front-to-back ratio at $E=7.1\times10^{-3}$. In (b) green and blue curves correspond to linear and nonlinear cases, respectively. The middle branches inside the bistability loops correspond to the unstable solutions. Arrows show the transitions between the stable branches. Transitions between the stationary states marked by red dots at $\Omega=-0.062$ and $\Omega=-0.11$ were studied by numerical simulations of Eq. (\ref{dynamic}).}
\end{figure}
We rewrite Eq. (\ref{Fourier transforms}) in the form
\begin{equation}\label{Fourier transforms 2}
\begin{array}{lcl}
p_{{\rm Si,}z}=\alpha_{\rm{Si}}( E_z^{(\rm{ex})}+A p_{{\rm Ag,}z}), \\
\alpha_{\rm Ag}^{-1}p_{{\rm Ag,}z}=E_z^{(\rm{ex})}\exp(ikd)+A p_{{\rm Si,}z}.
\end{array}
\end{equation}
Assuming that $\chi^{(3)} |{\bf E}^{(\rm{in})}|^2 \ll 1$ and $\nu/\om_0 \ll 1$, we decompose $\alpha^{-1}_{\rm{Ag}}(\om)$
in the vicinity of $\om_0$ and keep the first-order terms for the derivatives describing (actually small) broadening of the silver
particle polarization spectrum,
\begin{equation}
\label{polarizability}\alpha_{\rm{Ag}}^{-1} \approx \alpha_{\rm{Ag}}^{-1}(\om_0) + \left. \frac{d\alpha^{-1}_{\rm{Ag}}}{d\om} \right|_{\om=\om_0} \left(\Delta \om+i\frac{d}{dt}\right),
\end{equation}
where $\Delta \om$ is the frequency shift from the resonant value. Taking into account the instantaneous response of the Si nanoparticle and the relatively low strength of dipole-dipole interaction [$\left(R_{\rm{Si,Ag}}/d\right)^3<<1$], we set $\alpha_{\rm{Si}}=\alpha_{\rm{Si}}(\om_0)=\alpha_{\rm{Si}}^0$ and $A=A(\om_0)=A_0$. Having expressed ${\bf E}^{(in)}$ via ${\bf p}_{{ \rm Ag,}z}$, we
substitute Eq. (\ref{polarizability}) into Eq. (\ref{Fourier transforms 2}) and obtain the equations,
\begin{figure}[t]
\centerline{\rm{\resizebox{12cm}{!}{\includegraphics{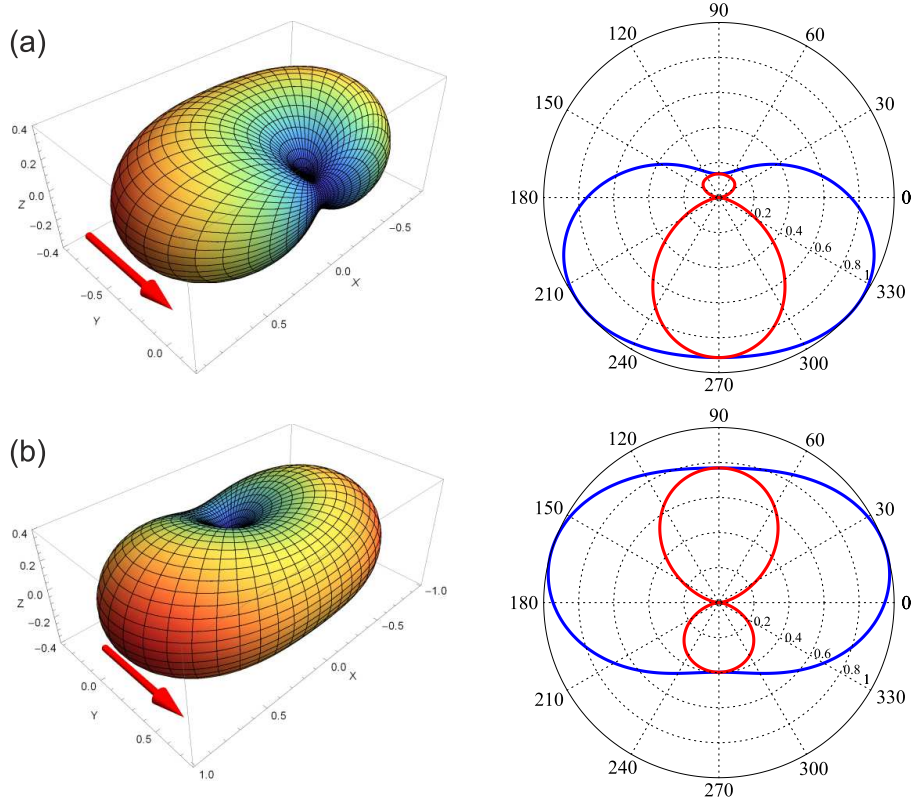}}}}
\caption{\label{fig:4} Normalized scattering patterns of the metal-dielectric nanoantenna for the stationary states denoted by red dots 1 (a) and 2 (b) in Fig.~\ref{fig:3}. Red and blue curves indicate $E$-plane and $H$-plane, respectively. Red arrows show the direction of plane wave incidence.}
\end{figure}
\begin{equation}\label{dynamic}
\begin{array}{lll}
P_{\rm{Si}}=\alpha_{\rm{Si}}^0[ E+A_0 P_{\rm{Ag}}],\\
i\frac{d P_{\rm{Ag}}}{d\tau}+\left(i\gamma+\Omega+|P_{\rm{Ag}}|^2 + \delta \Omega \right) P_{\rm{Ag}} = E \left[\exp(ik_0d)+\alpha_{\rm{Si}}^0 A_0\right],
\end{array}
\end{equation}
where $\delta\Omega=[3\veh^2 R_{\rm{Ag}}^3][2(\ve_{\infty}+2\veh)]^{-1}\alpha_{\rm{Si}}^0 A^2_0$ is the resonant frequency shift from $\om_0$ caused by the dipole-dipole interaction,
$P_{\rm Si,Ag}=p_{({\rm Si,Ag}),z}\sqrt{\chi^{(3)}}(\sqrt{2(\ve_\infty+2\veh)}\veh a^3)^{-1}$ and $E = -3 \veh \sqrt{\chi^{(3)}} E_z^{(\rm{ex})}[8(\ve_\infty+2\veh)^3]^{-1/2}$ are dimensionless slowly varying amplitudes of the particle dipole moments and the external electric field, respectively, $\gamma=\nu/(2\om_0)+(k_0 a)^3\veh(\ve_\infty+2\veh)^{-1}$ is in charge of thermal and radiation losses of the Ag particle, $k_0=\om_0/c\sqrt{\veh}$, $\Omega=(\om-\om_0)/\om_0$ and $\tau = \om_0 t$. Equation~(\ref{dynamic}) describes
temporal nonlinear dynamics of a hybrid Si-Ag dimer driven by the external plane wave with the frequency $\om\sim\om_0$.

Next, we consider the stationary solution of Eq. (\ref{dynamic}) which can be written as follows,
\begin{equation}\label{stationary solution}
\begin{array}{lll}
P_{\rm{Si}}=\alpha_{\rm{Si}}^0[ E+A_0 P_{\rm{Ag}}], \\
\left(i\gamma+\Omega+|P_{\rm{Ag}}|^2 + \delta \Omega \right) P_{\rm{Ag}} = E \left[\exp(ik_0d)+\alpha_{\rm{Si}}^0A_0\right].
\end{array}
\end{equation}
When $\Omega<-{\rm Re}\delta\Omega - \sqrt{3}\left( \gamma - {\rm Im}\delta\Omega \right)$, the particle polarizations $P_{\rm{Ag}}$ and $P_{\rm{Si}}$ become three-valued functions of $\Omega$ with two stable (lower and upper) and one unstable (middle) branches, as shown in Fig.~\ref{fig:3} (a). As a consequence, bistability, in this case, also appears in the frequency dependency of the nanoantenna front-to-back ratio (see Fig.~\ref{fig:3} (b)). One may see the significant contrast in the scattering patterns corresponding to the different stable branches. Being in the upper branch, the dimer scatters light predominantly in the forward direction or omnidirectionally, whereas the lower branch shows pronounced backscattering. Thus, having fixed $\Omega$ close to one of the thresholds of the bistability region, one can switch the system scattering pattern by varying the external field intensity.

\begin{figure}[t]
\centerline{\rm{\resizebox{13cm}{!}{\includegraphics{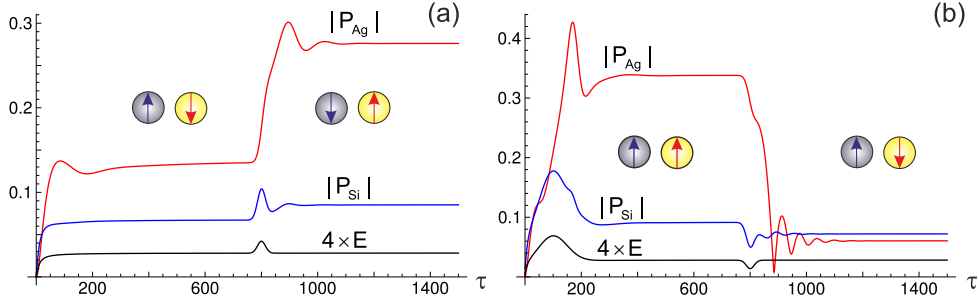}}}}
\caption{\label{fig:5} Temporal dependencies of $|P_{\rm{Ag}}|$ (red), $|P_{\rm{Si}}|$ (blue) and $E$ (black) when the signal pulse provokes a transition (a) from the back scattering state 1 to the forth scattering state 2 and (b) from the omnidirectional scattering state 3 to the back scattering state 4. Media 1 and Media 2 demonstrate dynamical behaviors of the system scattering pattern corresponding to (a) and (b), respectively.}
\end{figure}
To illustrate the dynamical control of the nanoantenna radiation, we solve Eq. (\ref{dynamic}) numerically
at $\Omega=-0.062$ ($\hbar\om\simeq2.94$~eV) and $\Omega=-0.11$ ($\hbar\om\simeq2.8$~eV) and zero initial conditions. The stable system states generated by
these frequencies are marked in Fig. \ref{fig:3} (b) by red dots, and the scattering patterns corresponding to $\Omega=-0.062$ are shown in Fig. \ref{fig:4}. To initiate the system transition from the state 1 to the state 2, we assume that the amplitude of the pump pulse varies slow
growing to the saturation level $E_0=7.1\times10^{-3}$. Then, when the system comes to the state 1, the Gaussian signal pulse appears,
provoking the system transition in the state 2, as shown in Fig. \ref{fig:5} (a).

To initiate the transition between the states 3 and 4, we add the Gaussian pulse centered around $\tau\sim100$ to the pump pulse slowly increasing to the level $E_0=7.1\times10^{-3}$ because the state 3 belongs to the upper branch of the bistability loop. Then, at $\tau\sim 800$ a signal pulse, being out of phase with a pump pulse, results in switching, as shown in Fig.~\ref{fig:5} (b).

In both cases a characteristic switching time is of $\tau\sim200$ or $\sim260$ fs, and the saturation amplitude of the external pulse $E_0=7.1\times10^{-3}$ corresponds to the intensity of $10.8$~MW/cm$^2$. However, such high illuminating powers can lead to thermal damage of the dimer. To estimate maximal pulse duration, we rely on the results of previous studies on the ablation thresholds for silver particles~\cite{Torres} and silicon films~\cite{Harrison} providing values about 3.96 J/cm$^2$ and 0.4 J/cm$^2$, respectively, for the picosecond regime of illumination. Taking into account amplification of the electric field inside the Ag nanoparticle due to surface plasmon resonance and the required intensity of $11$~MW/cm$^2$, we come to the maximal pulse durations of $0.9$~ns and $36$~ns corresponding to the Ag and the Si particles, which is much longer than the characteristic switching time. Thus, the proposed metal-dielectric nanoantenna is fully suitable for ultra fast all-optical switching. Furthermore, it looks to be a competitive alternative with optical switches based on semiconductor microcavities whose minimal switching time is of several picoseconds \cite{menon_nat_phot_10}.

Apparently, when the external electric field is polarized along y-axis, in the linear regime one can excite in the dimer a longitudinal dipole mode at $\Omega=0$ or linear quadrupole modes with positive and negative relative phase shifts at $\Omega>0$ and $\Omega<0$, respectively \cite{Shegai,Alivisatos}. Consequently, a bistable response of the silver nanoparticle will result in the dipole-quadrupole and quadrupole-quadrupole dynamical transitions of the system scattering pattern in the manner similar to that described above.

In practice Si-Ag heterodimers can be obtained through a recently suggested combination of top-down fabrication and template-guided self-assembly \cite{Boriskina} which allows a precise and controllable vertical and horizontal positioning of the plasmonic elements relative to dielectric spheres. Atomic force microscopy nanomanipulation is another useful approach for construction of such hybrid systems \cite{Benson_NL,Ratchford,Benson_Nature}.

\section{Conclusions}

We have predicted a bistable response of a hybrid Si-Ag nanoantenna and revealed that its radiation pattern can be reversed by varying the external intensity. This effect originates from the phase shift between the dipole moments of the nanoparticles caused by differences in polarizabilities between Si and Ag nanoparticles induced by intrinsic nonlinearity of a silver particle and different dispersions and sizes of particles. A characteristic switching time of 260 fs along with relatively low required intensities of $11$~MW/cm$^2$ can be useful for ultrafast logical hybrid nanophotonic-plasmonic devices and circuitry.

\section*{Acknowledgements}

The authors are indebted to A.A. Zharov, A.E. Miroshnichenko, and I.S. Maximov for useful comments, to P.A. Belov for stimulating discussions, and to the Ministry of Education and Science of Russian Federation and the Australian Research Council for a financial support.

\section*{References}
\bibliographystyle{unsrt}
\bibliography{References}

\end{document}